\documentclass[aps,prx,superscriptaddress,amsmath,amssymb,twocolumn,floatfix,reprint]{revtex4-2}

\usepackage[utf8]{inputenc}
\usepackage[T1]{fontenc}
\usepackage{graphicx}
\usepackage{braket}
\usepackage{multirow}
\usepackage{makecell}
\usepackage[dvipsnames]{xcolor}
\usepackage{xr}
\usepackage{enumitem}  
\usepackage{lineno}

\usepackage[colorlinks=true,
            urlcolor=blue,
            linkcolor=blue,
            citecolor=blue,
            pdftex]{hyperref}
\usepackage[capitalise]{cleveref}

\begin{document}
\title{Scaling Laws for Neural-Network Quantum States}

\author{Riccardo Rende}
\email{rrende@flatironinstitute.org}
\affiliation{Center for Computational Quantum Physics, Flatiron Institute, 162 5th Avenue, New York, NY 10010}

\author{Alessandro Sinibaldi}
\email{alessandro.sinibaldi@epfl.ch}
\affiliation{Institute of Physics, \'{E}cole Polytechnique F\'{e}d\'{e}rale de Lausanne (EPFL), CH-1015 Lausanne, Switzerland}

\author{Luciano Loris Viteritti}
\email{luciano.viteritti@epfl.ch}
\affiliation{Institute of Physics, \'{E}cole Polytechnique F\'{e}d\'{e}rale de Lausanne (EPFL), CH-1015 Lausanne, Switzerland}

\author{Roeland Wiersema}
\affiliation{Center for Computational Quantum Physics, Flatiron Institute, 162 5th Avenue, New York, NY 10010}

\author{Antoine Georges}
\affiliation{Center for Computational Quantum Physics, Flatiron Institute, 162 5th Avenue, New York, NY 10010}
\affiliation{Coll{\`e}ge de France, 11 place Marcelin Berthelot, 75005 Paris, France}
\affiliation{CPHT, CNRS, {\'E}cole Polytechnique, IP Paris, F-91128 Palaiseau, France}
\affiliation{DQMP, Universit{\'e} de Gen{\`e}ve, 24 quai Ernest Ansermet, CH-1211 Gen{\`e}ve, Suisse}

\author{Giuseppe Carleo}
\affiliation{Institute of Physics, \'{E}cole Polytechnique F\'{e}d\'{e}rale de Lausanne (EPFL), CH-1015 Lausanne, Switzerland}

\date{\today}
\begin{abstract}
Scaling laws, the power-law relations between loss, architecture size, and compute observed in modern neural networks, offer a quantitative way to characterize the complexity of a learning problem, with the exponent governing the decay of the loss reflecting how rapidly additional resources translate into improved accuracy, and thus how hard the target is to learn. Whether an analogous framework can characterize the complexity of physical problems remains open. We address this question for Neural-Network Quantum States, a leading variational approach for strongly correlated quantum many-body systems. Using transformer wave functions to approximate ground states of the $J_1$-$J_2$ Heisenberg model on triangular and square lattices with up to $20\times 20$ sites, we find that the $V$-score, a measure of accuracy of a variational state, decays as a power law in training compute. Under an appropriate rescaling of compute, results for different system sizes collapse onto a single curve, analogous to scaling collapse in critical phenomena. The resulting power law is, to a good approximation, independent of the number of sites, showing that the transformer Ansatz is size-consistent for the systems considered. The exponent decreases systematically with frustration, identifying it as a quantitative measure of representational difficulty of the ground state and establishing scaling laws as a general framework for benchmarking variational ans\"{a}tze.
\end{abstract}

\maketitle

\section{Introduction}
A robust empirical observation in modern deep learning is that the test loss of large neural networks decreases as a power law with architecture size, training dataset size, and training compute. First quantified for autoregressive transformer language models~\cite{kaplan2020scaling} and later refined~\cite{hoffmann2022chinchilla}, this behavior has since been observed for vision  transformers~\cite{zhai2022scalingvit,dehghani2023vit22b,alabdulmohsin2023sovit} and across a range of other domains~\cite{henighan2020scaling,hilton2301scaling,neumann2022scaling}. Its central implication is that increasing training resources by a fixed factor leads to a predictable reduction in test loss, with no sharp threshold or saturation over many orders of magnitude~\cite{hoffmann2022chinchilla}. The main practical value of scaling laws is predictive: a small number of modest-scale training runs can be used to estimate the loss of much larger architectures. This makes it possible to choose in advance which architecture to train and how much compute to allocate, rather than relying on costly trial and error~\cite{kaplan2020scaling,hoffmann2022chinchilla}.

Recent theoretical work has begun to explain the origin of such power-law behavior in simplified settings. These studies show that scaling exponents depend systematically on the architecture, the data distribution, and the task, and that they change in a controlled way when any of these ingredients is modified~\cite{cagnetta2026,bahri2024explaining}. This perspective suggests that scaling laws can be used not only to predict model performance, but also to extract information about the underlying data distribution. At present, however, these theoretical frameworks remain limited to simplified settings and do not fully capture the architectural details of large transformer models with millions or billions of parameters~\cite{dehghani2023vit22b}. As a result, the evidence for scaling laws in state-of-the-art architectures remains primarily empirical.

Transformer architectures, originally developed for natural-language processing and the setting in which scaling laws were first established, have also become powerful tools across the physical and natural sciences. Applications now range from quantum many-body physics~\cite{viteritti2023transformer,sobral2025,nutakki2026vit,raikos2026vit,yamazaki2026vit,cao2024vit,viteritti2026disorder} and electronic-structure theory to chemistry~\cite{vonglehn2023,shang2025solving}, materials discovery~\cite{mswahili2024,pyzerknapp2025foundation}, and protein-structure prediction~\cite{jumper2021highly,moussad2023}. Whether the scaling laws that govern transformers in language and vision carry over to these scientific settings, and whether they retain their predictive and diagnostic value, is largely unexplored. 

In this work, we address this question for quantum many-body physics, within the framework of Neural-Network Quantum States (NQS)~\cite{carleo2017solving}. This variational approach was introduced to circumvent the exponential growth of the Hilbert space by representing the wave function with an artificial neural network and optimizing its variational parameters within the Variational Monte Carlo framework~\cite{becca2017}. 
After the original work of~\citet{carleo2017solving}, which focused on relatively simple fully connected neural networks, subsequent studies explored a broad range of neural-network architectures for representing variational states~\cite{lange2024review}, including convolutional~\cite{choo2019twodimensional,chen2024empowering,chen2025convolutional,hendry2025grassmann}, recurrent~\cite{rnn2020,moss2025rnn}, and autoregressive architectures~\cite{sharir2020deep}. More recently, transformer-based architectures~\cite{viteritti2023transformer,sprague2024variational,chen2025convolutional,rende2025foundation} have delivered some of the most accurate variational results across a wide range of quantum many-body problems, including two-dimensional frustrated spin models~\cite{viteritti2023transformer,viteritti2025prb,chen2025convolutional,fan2026, viteritti2026spatialattention}, fermionic lattice systems~\cite{gu2025,viteritti2026,rende2026ancilla}, Rydberg atom arrays~\cite{sprague2024variational}, quantum impurity models~\cite{kim2024impurity}, and electronic-structure calculations~\cite{vonglehn2023,shang2025qiankunnet,zhao2023scaling}.

Since transformer-based architectures underlie both the scaling laws of language modeling and state-of-the-art NQS ans\"{a}tze, it is natural to ask whether analogous power-law behavior also governs their performance in this very different setting. Two recent results on the scaling behavior of NQS are particularly relevant in this context. On the empirical side, a scaling study of autoregressive NQS applied to electronic-structure problems~\cite{knitter2024llmscaling} reported Chinchilla-type behavior~\cite{hoffmann2022chinchilla}. On the theoretical side, Ref.~\cite{lu2026information} derived information-theoretic scaling laws for autoregressive NQS.

\begin{figure*}[t]
  \centering
  \includegraphics[width=2.0\columnwidth]{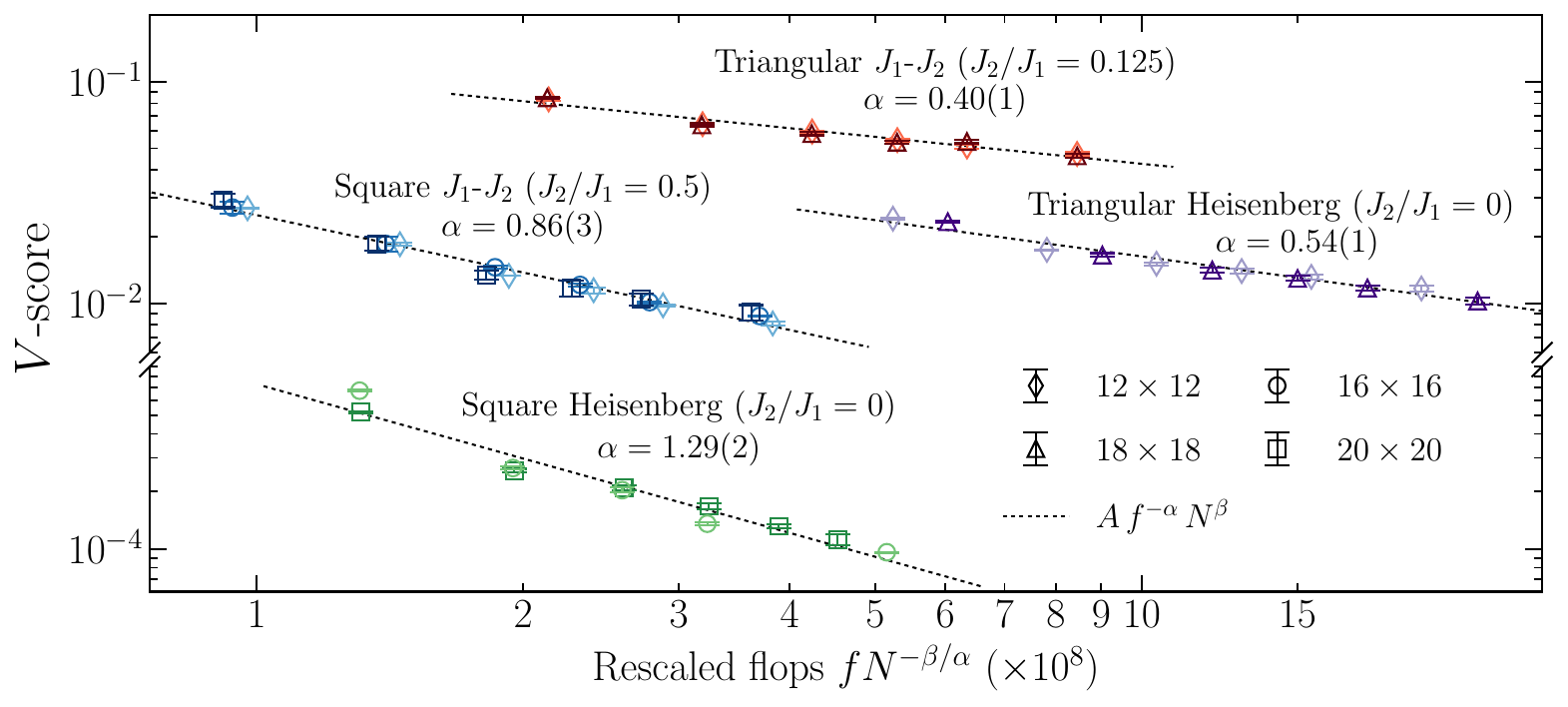}
  \caption{
  Scaling collapse of the $V$-score for transformer wave function on four two-dimensional spin Hamiltonians: square Heisenberg, triangular Heisenberg, square $J_1$-$J_2$ Heisenberg, and triangular $J_1$-$J_2$ Heisenberg model (see \cref{sec:ham_model}).
  For each model, $V$-score data obtained at linear system sizes $L = 12, 16, 18, 20$ are plotted against the rescaled compute $f N^{-\beta/\alpha}$, where $f$ is the training compute in FLOPs and $\alpha$, $\beta$ are the exponents of the scaling Ansatz of \cref{eq:scaling}.
  After rescaling, data from all system sizes collapse onto a single power law $V\text{-score} = A\, f^{-\alpha}\, N^{\beta}$ (dashed black lines), with the fitted exponent $\alpha$ indicated next to each data set. The exponent varies systematically across Hamiltonians, decreasing from the unfrustrated square Heisenberg model [$\alpha = 1.29(2)$] to the strongly frustrated triangular $J_1$-$J_2$ Heisenberg model [$\alpha = 0.40(1)$]. Fitted values of $\alpha$, $\beta$, and $A$ for all four models are listed in \cref{tab:exponents}.}
  \label{fig:main_results}
\end{figure*}

To address this question, we study the scaling behaviour of transformer-based NQS in finding the ground state of two-dimensional frustrated quantum spin systems. In particular, we focus on the $J_1$-$J_2$ Heisenberg model, on both the square and triangular geometries. We consider $L \times L$ clusters with periodic boundary conditions with linear sizes up to $L=20$.
As a measure of accuracy of the variational state, we use the $V$-score metric introduced in Ref.~\cite{wu2024vscore}, defined as
\begin{equation}
    V\text{-score} = N \frac{\langle{\hat{H}^2}\rangle - \langle \hat{H} \rangle^2}{\langle \hat{H} \rangle^2} \ ,
    \label{eq:vscore}
\end{equation}
where $N=L^2$ is the number of sites. The $V$-score combines the energy and variance of the variational state into a single dimensionless quantity that correlates with the relative energy error. It is positive definite and vanishes for the exact ground state, making it analogous to a loss function in deep-learning settings~\cite{wu2024vscore}. Here, we study its scaling with training compute, measured by the number of floating-point operations (FLOPs) $f$ required to perform forward evaluations of the variational state during the ground-state optimization. \\
Our results establish three main findings:
\begin{enumerate}[label=(\roman*)]
\item The $V$-score decays as a power law in training compute, and the exponent of this decay provides a single-number measure of how efficiently a transformer wave function learns a given target ground state.
\item For each Hamiltonian, after rescaling compute, data from all system sizes collapse onto a single curve. Within numerical uncertainty, the collapse exponent is independent of system size and architectural details, which affect only the prefactor, as observed in language and vision models~\cite{hoffmann2022chinchilla,pearce2024reconciling,li2025farseer}. This demonstrates that the transformer Ansatz is \textit{size consistent}, allowing the compute required to reach a target accuracy on large systems to be estimated from smaller-system simulations.
\item The exponent of the power law decreases with model frustration, identifying it as a quantitative measure of the representational complexity of the target ground state.
\end{enumerate}

Together, these results on scaling laws for NQS provide not only a diagnostic of ground state's Hamiltonian complexity, but also a practical tool for guiding the design and selection of variational architectures.

\section{The \texorpdfstring{$J_1$-$J_2$}{J1-J2} Heisenberg Model}\label{sec:ham_model}
We focus on the spin-$1/2$ $J_1$-$J_2$ Heisenberg model, a paradigmatic setting for frustrated quantum magnetism. The Hamiltonian is
\begin{equation}
\hat{H} = J_1 \sum_{\langle i,j \rangle} \hat{\boldsymbol{S}}_i \cdot \hat{\boldsymbol{S}}_j + J_2 \sum_{\langle\langle i,j \rangle\rangle} \hat{\boldsymbol{S}}_i \cdot \hat{\boldsymbol{S}}_j,
\label{eq:hamiltonian}
\end{equation}
where $\hat{\boldsymbol{S}}_i$ is the spin-$1/2$ operator at site $i$, and $\langle i,j \rangle$ and $\langle\langle i,j \rangle\rangle$ denote nearest- and next-nearest-neighbour pairs, respectively. The sites form a two-dimensional $L \times L$ lattice with periodic boundary conditions. \\
We consider four representative cases:
\begin{itemize}[label=\scriptsize$\bullet$]
    \item \textit{Square Heisenberg ($J_2/J_1=0)$}: the ground state exhibits Néel order. This model is free of the sign problem and can therefore be solved using Quantum Monte Carlo methods~\cite{sandvik1997, sandvik2026high, calandra1998};
    \item \textit{Square $J_1$-$J_2$ Heisenberg at $J_2/J_1=0.5$}: the competing nearest- and next-nearest-neighbour interactions strongly frustrate the system, and the ground state is expected to be a gapless quantum spin liquid~\cite{becca2013, nomura2021, chen2024empowering};
    \item \textit{Triangular Heisenberg ($J_2/J_1=0)$}: geometric frustration leads to a ground state with $120^\circ$ Néel antiferromagnetic order~\cite{capriotti1999, white2007, viteritti2026spatialattention};
    \item \textit{Triangular $J_1$-$J_2$ Heisenberg at $J_2/J_1=0.125$}: the additional competing interaction further frustrates the system, and the ground state is expected to be a gapless quantum spin liquid~\cite{kaneko2014gapless, iqbal2016, chen2024empowering}.
\end{itemize}

Together, these Hamiltonians provide a controlled set of target ground states spanning different levels of  representational difficulty, allowing us to compare scaling behavior across physically distinct regimes.

\section{Critical Behaviour of the \texorpdfstring{$V$}{V}-score}
Inspired by critical phenomena, where power-law behavior and scaling collapse emerge near critical points~\cite{wilson1974renormalization,wilson1975renormalization,goldenfeld1972,cardy1996scaling}, and by empirical evidence of scaling laws in machine learning~\cite{kaplan2020scaling,hoffmann2022chinchilla}, we propose a scaling form for the accuracy of Neural-Network Quantum States. Specifically, we describe the dependence of the $V$-score [see \cref{eq:vscore}] on the training compute $f$ and on the number of sites $N$ as
\begin{equation}
    V\text{-score} = A\, f^{-\alpha}\, N^{\beta}, 
    \label{eq:scaling}
\end{equation}
where $\alpha,\beta>0$ are scaling exponents and $A$ is a non-universal prefactor.

In the following, we study the power-law decay of the $V$-score with training compute across different system sizes and use the extracted exponents to characterize the scaling behavior of transformer-based variational states.

\begin{figure*}[t]
  \centering
  \includegraphics[width=2\columnwidth]{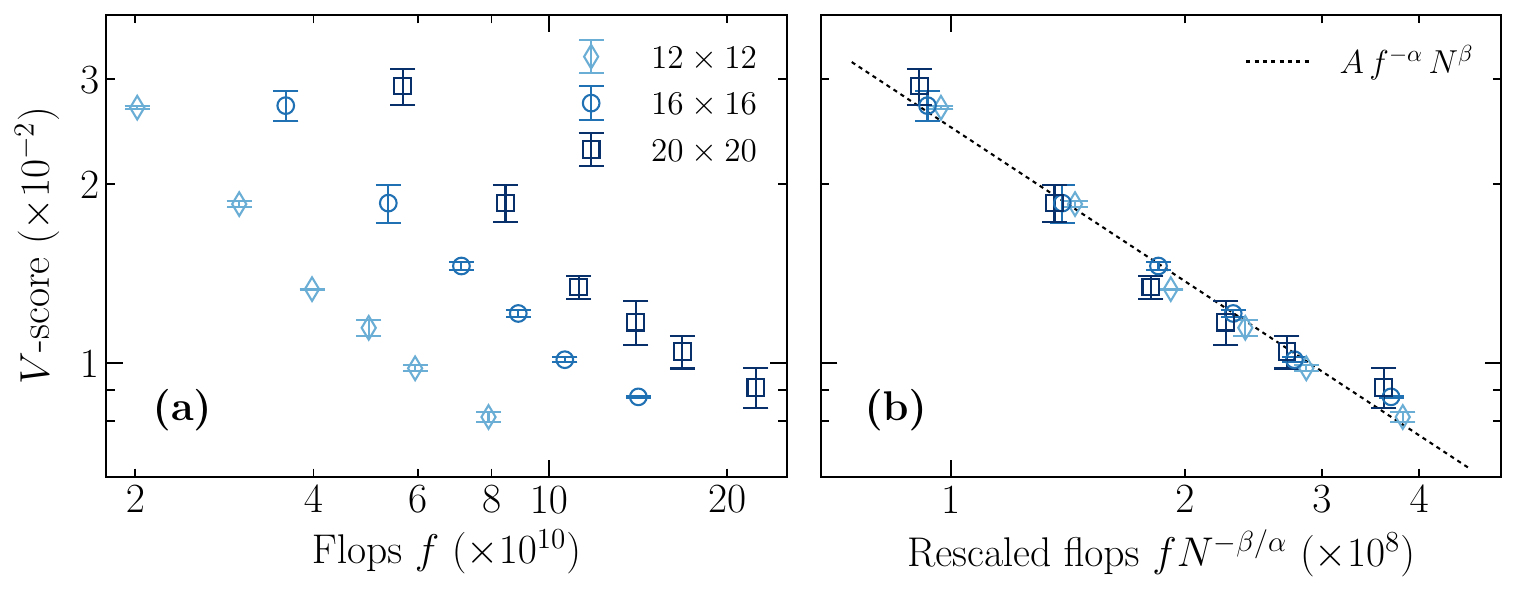}
  \caption{\label{fig:collapse} Illustration of the scaling collapse procedure for the square $J_1$-$J_2$ Heisenberg model. \textbf{Panel (a):} $V$-score as a function of the training compute $f$ (FLOPs) for linear system sizes $L = 12, 16, 20$. \textbf{Panel (b):} The same data plotted against the rescaled compute $f N^{-\beta/\alpha}$, with $\alpha$ and $\beta$ taken from \cref{tab:exponents}, collapse onto a single curve well described by the power law $V\text{-score} = A\, f^{-\alpha} N^\beta$ (dashed black line) [see \cref{eq:VPL}].}
\end{figure*}

Our main results are summarized in \cref{fig:main_results}, which shows the decay of the $V$-score for the four representative two-dimensional spin Hamiltonians introduced above. For each Hamiltonian, we train transformer wave functions on clusters ranging from $12 \times 12$ to $20 \times 20$, varying the number of layers to systematically increase the training compute $f$. \cref{fig:main_results} shows the $V$-score as a function of the rescaled training compute $fN^{-\beta/\alpha}$. For each Hamiltonian, the exponents $\alpha$ and $\beta$ are obtained through a collapse procedure, chosen so that data from different system sizes fall onto a single curve (see below). This rescaling removes the explicit system-size dependence of the $V$-score and reveals a universal scaling curve for each model. Although the scaling form contains two exponents, $\alpha$ and $\beta$, we find that they are related in the regime studied here, refer to \cref{sec:size_consistency} for details. As a result, the variational difficulty of finding the ground state of a given Hamiltonian with a transformer architecture can be quantified by a single number: the exponent $\alpha$ of the collapsed power law. This exponent measures, in a size-independent way, how efficiently additional compute improves the variational accuracy for a given target Hamiltonian.

Remarkably, $\alpha$ is robust across architectural choices, and optimization details, see \cref{app:validity} and \ref{sec:ansatz}. These choices affect only the non-universal prefactor $A$, in line with observations from scaling laws in language and vision models~\cite{hoffmann2022chinchilla,pearce2024reconciling,li2025farseer}.

The collapse procedure is illustrated explicitly in \cref{fig:collapse} for the square $J_1$-$J_2$ Heisenberg model. The three curves show a clear dependence on system size when plotted as a function of $f$ [see panel (a)]. Applying the scaling form in \cref{eq:scaling}, the data collapse onto a single curve when plotted as a function of the rescaled compute $fN^{-\beta/\alpha}$ [see panel (b)].

Finally, \cref{tab:exponents} reports the fitting parameters $\alpha$, $\beta$ and $A$ obtained from the collapse procedure for the four models. The values of $\alpha$ vary systematically across the Hamiltonians. The unfrustrated square Heisenberg model exhibits the steepest decay, with $\alpha = 1.29(2)$, whereas the triangular $J_1$-$J_2$ Heisenberg model exhibits the shallowest decay, with $\alpha = 0.40(1)$. The square $J_1$-$J_2$ Heisenberg model, with $\alpha = 0.86(3)$, and the triangular Heisenberg model, with $\alpha = 0.54(1)$, lie between these two limits.

This ordering reflects the intrinsic physical complexity of the corresponding ground states: frustration, whether geometric as in the triangular lattice or induced by competing exchange interactions, reduces $\alpha$ and therefore slows the rate at which additional compute improves variational accuracy. For the transformer architecture considered here, the triangular Heisenberg model therefore appears more difficult to optimize than the square $J_1$-$J_2$ Heisenberg model. While the connection between frustration and the representational complexity of the ground state is qualitatively expected, the exponent $\alpha$ provides a quantitative and systematic measure of this complexity within a neural-network variational Ansatz.

\begin{table}[t]
  \centering
  \begin{tabular}{|c|c|c|c|}
    \hline\hline
    \textbf{Hamiltonian} & $\boldsymbol{\alpha}$ & $\boldsymbol\beta$ & $\boldsymbol{\log A}$ \\
    \hline
    Square Heisenberg $(J_2/J_1=0)$     & $1.29(2)$ & $1.31(3)$ & $16.5(5)$ \\
    \hline
    Square $J_{1}$--$J_{2}$ $(J_2/J_1=0.5)$    & $0.86(3)$ & $0.93(4)$ & $12.2(5)$ \\
    \hline
    Triangular Heisenberg $(J_2/J_1=0)$ & $0.54(1)$ & $0.46(2)$ & $7.2(3)$  \\
    \hline
    Triangular $J_{1}$--$J_{2}$ $(J_2/J_1=0.125)$ & $0.40(1)$ & $0.42(2)$ & $5.2(2)$  \\
    \hline\hline
  \end{tabular}

  \caption{\label{tab:exponents}
    Fitted exponents and prefactors of the scaling
    Ansatz~\eqref{eq:scaling} for the four lattice models studied.
    The error is estimated via resampling technique with gaussian noise.}
\end{table}

\subsection{Size Consistency}
\label{sec:size_consistency}

The general scaling form of the $V$-score in \cref{eq:scaling} can be specialized to the neural-network architecture used in this work. As we show below, this step is important for interpreting the scaling exponents and for understanding their dependence on system size.

For the transformer architecture considered here, the computational cost $f$ contains two contributions: one linear and one quadratic in the number of sites $N$. In the typical regime in which NQS calculations are performed, the leading contribution is the linear term; additional details are given in \cref{app:flops}. We therefore approximate $f \approx a N$, where $a$ is an architecture-dependent constant. Substituting this relation into the scaling Ansatz of \cref{eq:scaling} gives
\begin{equation}
    V\text{-score} \approx a^{-\alpha} N^{-\alpha+\beta} \ .
    \label{eq:VPL}
\end{equation}

The fitted exponents reported in \cref{tab:exponents} satisfy, within numerical error, $\alpha \approx \beta$ for all Hamiltonians considered here. The explicit dependence on $N$ therefore cancels out, and as a result the scaling laws are size independent.
This simple argument shows that the $V$-score is independent of the number of sites. We refer to this property as the \emph{size consistency} of the variational state~\cite{viteritti2026spatialattention}.

From the computational side, size consistency has a direct practical implication. Since the exponent $\alpha$ is independent of $N$, the compute required to reach a target $V$-score on a large system can be estimated from smaller-system simulations. In practice, a small number of training trajectories at modest $N$ is sufficient to determine the scaling parameters, after which the scaling law can be extrapolated to larger system sizes without additional fitting, provided the system remains in the regime where the scaling laws hold; see \cref{app:validity} for further details.

\subsection{Robustness of Scaling Law}
\label{sec:ansatz}

\begin{figure}[t]
  \centering
  \includegraphics[width=1\columnwidth]{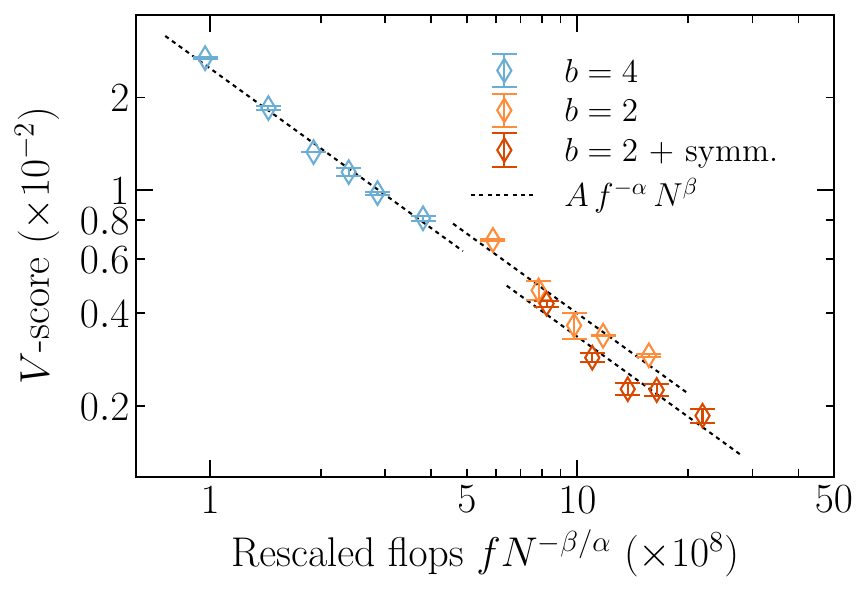}
\caption{\label{fig:ansatz} $V$-score as a function of the rescaled compute $f N^{-\beta/\alpha}$ and corresponding power law fit for different architectural choices: the transformer architecture with linear patch size $b=4$ as in ~\cref{fig:main_results}, the architecture with a smaller $b=2$, and the architecture with $b=2$ and augmented with translation symmetries.
The system considered here is the square $J_1$-$J_2$ Heisenberg model with $L=12$.}
\end{figure}

Another key result of our analysis is that the exponent $\alpha$ is robust to architectural choices in the transformer. 
\cref{fig:ansatz} shows the $V$-score curves for two distinct architectural variants  (see~\cref{app:vit} for details on the hyperparameters of the neural network).
First, we consider a linear patch size $b=2$, namely smaller compared to $b=4$ used for the main results in \cref{fig:main_results}.
Second, we enforce translational symmetry on the wave function with $b=2$ via quantum-number projection~\cite{nomura2021helping,schmittsymm,viteritti2023transformer,viteritti2025prb}, which yields a superposition that is not in general representable as a single transformer. 
Both modifications increase the number of FLOPs and reduce the $V$-score, while leaving the scaling behavior unchanged.
The data lie well on the power law fit with same exponents $\alpha$ and $\beta$ from \cref{tab:exponents}, demonstrating that the scaling exponent is essentially invariant under architectural changes~\cite{hoffmann2022chinchilla,pearce2024reconciling,li2025farseer}.
The prefactor $A$, by contrast, is architecture-dependent and does not constitute a universal property of the Hamiltonian.

Based on the data in~\cref{fig:ansatz}, we observe that to obtain the lowest $V$-score achieved by the model with $b=2$ and the symmetries using the unsymmetrized Ansatz with $b=4$ would require roughly $40$ layers.
Beyond this depth, however, training becomes more challenging and the scaling laws no longer apply, as discussed in~\cref{app:validity}.
We therefore conclude that incorporating symmetries directly into the variational wave function is important for achieving low $V$-scores, rather than relying on very deep unsymmetrized wave functions.

\section{Conclusions}
We have shown that the empirical scaling laws observed in large language and vision transformers also apply to neural-network variational representations of quantum many-body wave functions. For the four two-dimensional spin Hamiltonians considered, the $V$-score of a transformer Ansatz decays as a power law in training compute. This places variational quantum many-body methods within the same quantitative framework used in deep learning, and allows architecture and compute choices to be made on a quantitative basis.

Two properties of this scaling law are worth noting. 
First, for the transformer architecture the power law is, to a good approximation, effectively independent of the number of sites in the system. This is what makes the transformer a size-consistent architecture for the prototypical Hamiltonians of quantum magnets. Second, the power law exponent $\alpha$ decreases from unfrustrated to frustrated models, and from square to triangular geometries. Thus, it provides a single number that can capture the degree of complexity of the target wave function.

A natural next step is to relate $\alpha$ to established measures of quantum state complexity: the entanglement entropy~\cite{hastings2010measuring,passetti2023can,comment}, the bond dimension of tensor-network representations~\cite{Or_s_2014}, non-stabiliserness monotones~\cite{turkeshi2025magic,tarabunga2023many,leone2022stabilizer,sinibaldimagic}, and other complexity measures~\cite{Mendes_Santos_2024}. 
Relating $\alpha$ to one of these quantities, and checking whether it is preserved under changes of basis and architecture, would connect the empirical scaling laws reported here to the structure of the target state. Extending the analysis to fermionic~\cite{xu2024,roth2025} and \emph{ab initio} electronic problems~\cite{spencer2021ab,nys2024ab,vonglehn2023} is a natural direction for future work. 

Since the exponent $\alpha$ quantifies the representational complexity of the target ground state, studying how this exponent changes across a phase transition is a natural direction for future works~\cite{rendefinetuning2024}. In particular, it could provide an additional diagnostic for detecting phase transitions: near criticality, where correlations become long ranged and the ground state is expected to be more harder to capture, one may expect a slower improvement of the $V$-score with compute, corresponding to a smaller value of the exponent.

\section*{Methods}
\subsection{Vision Transformer Wave Function and Optimization Details}
\label{app:vit}
In this Section we briefly summarize the variational Ansatz used in this work, referring to Ref.~\cite{viteritti2026spatialattention} for a detailed description.
The wave function is parametrized by a vision transformer architecture~\cite{viteritti2023transformer,viteritti2025prb} that maps a spin configuration $\boldsymbol{\sigma} = (\sigma_1, \dots, \sigma_N)$, with $\sigma_i = \pm 1$, to a complex amplitude $\Psi_\theta(\boldsymbol{\sigma})$ where $\theta$ are the variational parameters. The input configuration is partitioned into $n = L^2/b^2$ non-overlapping patches of size $b \times b$, each linearly embedded into a vector $\boldsymbol{x}_i \in \mathbb{R}^d$ through a learnable transformation. The resulting sequence $(\boldsymbol{x}_1, \dots, \boldsymbol{x}_n)$ is then processed by a stack of $n_l$ transformer layers, each implementing a Multi-Head Attention mechanism followed by a feed-forward block.
The central ingredient of the architecture is the \emph{Spatial Attention} mechanism introduced in Ref.~\cite{viteritti2026spatialattention}, in which the standard attention weights are reweighted by a distance-dependent kernel:
\begin{equation}
A_i = \sum_{j=1}^{n} \frac{e^{-\gamma\, d(i,j)}}{\sum_{j'=1}^{n} e^{-\gamma\, d(i,j')}}\, \alpha_{ij}\, V x_j,
\label{eq:spatial_attention}
\end{equation}
where $\alpha_{ij}$ are trainable attention coefficients, $V$ is the value matrix, and $d(i,j)$ is the Euclidean distance between patches $i$ and $j$. The inverse length scale $\gamma > 0$ is learned independently for each attention head and layer, and controls the characteristic spatial range over which patches contribute to the updated representation. This learnable geometric prior preserves global connectivity, with standard attention recovered in the limit $\gamma \to 0$, and it constitute the key ingredient to optimize large lattices~\cite{viteritti2026spatialattention}.

The specific transformer architecture used in this work comprises $h=12$ attention heads and embedding dimension $d=72$, with the number of layers varying from $n_l=2$ to $n_l=8$ unless otherwise specified. The linear size of the patch $b$ is chosen to be $b=2$ or $b=4$ for square and $b=3$ for triangular geometry.
A detailed description of the role of the different hyperparameters of the architecture is provided in Ref.~\cite{viteritti2025prb}. The variational parameters are optimized using the MARCH optimizer~\cite{gu2025} with a learning rate $\eta = 5 \times 10^{-3}$ and a cosine decay schedule over $N_\mathrm{opt} = 10\,000$ optimization steps, using the Stochastic Reconfiguration (SR)~\cite{sorella1998} formulation developed in Refs.~\cite{chen2024empowering,rende2024stochastic} which is optimal when the number of variational parameters is bigger than the number of Monte Carlo samples.
The SR update employs a diagonal shift regularization of $\lambda = 10^{-4}$. At each optimization step, the energy and its gradient are estimated via Variational Monte Carlo~\cite{becca2017} using $M= 16\,384$, unless otherwise specified, samples drawn from $|\Psi_\theta(\boldsymbol{\sigma})|^2$.

\subsection{Estimation of \texorpdfstring{$V$}{V}-score Error Bars}
\label{app:error_bars}
The $V$-score reported in the main text is affected by two distinct sources of uncertainty. The first is the statistical error inherent in the Variational Monte Carlo estimates of the mean energy $\langle \hat{H} \rangle$ and the energy variance $\mathrm{Var}(\hat{H})$, which together define the $V$-score [see \cref{eq:vscore}]. For a given variational state, this error is controlled by the number of Monte Carlo samples $M$ drawn from $|\Psi_\theta(\boldsymbol{\sigma})|^2$ and can be made arbitrarily small by increasing $M$. In our simulations, we use approximately $1.6$ million samples, which is sufficient to render this contribution negligible compared to the second source of error described below. We therefore do not include it in the error bars reported in this work.

The dominant contribution to the uncertainty originates from the dependence of the optimization outcome on the random seed. Even when the architecture, the optimizer, and all hyperparameters are kept fixed, different seeds produce slightly different final variational states, and hence different $V$-scores.
To quantify this variability, for each combination of Hamiltonian, number of sites $N$, and architectural configuration, we perform three independent optimization runs differing only in the random seed used to initialize the variational parameters and to draw the Monte Carlo samples. Among the three runs, we discard the worst-performing one and retain the two runs with the lowest final $V$-score. The reported $V$-score is taken to be the smaller of the two, while the associated error bar is estimated as the standard deviation between the $V$-scores of these two best runs. This procedure provides a reasonable estimate of the spread induced by seed variability, while reducing the sensitivity to outlier runs that fail to converge to a competitive minimum. 

\begin{figure}[t]
  \centering
  \includegraphics[width=1\columnwidth]{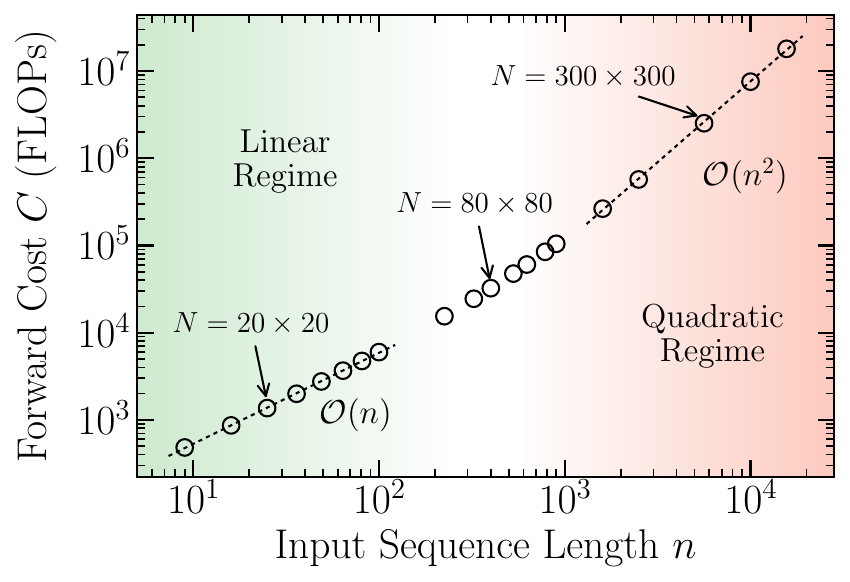}
\caption{\label{fig:complexity} 
Cost $C$ of a single forward evaluation of the transformer wave function on one spin configuration measured in FLOPs as a function of the length of the input sequence $n=L^2/b^2$.
The transformer architecture considered for this analysis has linear patch size $b=4$, number of layers $n_l=8$ and embedding dimension $d=72$.
A linear scaling regime of the cost $C$ with $n$ (highlighted in green) and a quadratic scaling regime (highlighted in red) are identifiable.}
\end{figure}

Moreover, we note that the standard Monte Carlo estimator of the energy variance $\mathrm{Var}(\hat{H})$ entering the $V$-score [see Eq.~(\ref{eq:vscore})] is biased in the presence of nodes in the wave function, due to the support-mismatch pathology analyzed in Refs.~\cite{sinibaldi2023unbiasing,wan2026blurred}. This bias originates from configurations $\boldsymbol{\sigma}$ with $\Psi_{\theta}(\boldsymbol{\sigma}) = 0$ but $\langle \boldsymbol{\sigma} | \hat{H} | \Psi_{\theta} \rangle \neq 0$, which are never visited by Monte Carlo sampling from $|\Psi_{\theta}(\boldsymbol{\sigma})|^2$ and therefore do not contribute to the stochastic estimator, even in the infinite-sample limit. In principle, this bias could be removed by employing the blurred sampling scheme of Ref.~\cite{wan2026blurred}. We have verified on representative cases that, for the system sizes and Hamiltonians considered in this work, the resulting corrections do not modify the reported $V$-scores within the estimated error bars.

\subsection{FLOPs Computation}
\label{app:flops}

In this section we describe how the training compute $f$, used as the horizontal axis in all the plots, is estimated for the vision transformer wave function~\cite{viteritti2023transformer,viteritti2025prb,viteritti2026spatialattention}.

We first measure compute in floating-point operations (FLOPs) required to perform a single forward evaluation of the variational state on one spin configuration $\boldsymbol{\sigma}$. For a transformer with $n_l$ layers, $n = L^2/b^2$ input length (number of patches), and embedding dimension $d$, the cost of one forward pass is dominated by two contributions: the linear projections inside each attention and feed-forward block, which scale as $\mathcal{O}(n\, d^2)$ per layer, and the attention operation itself, which scales as $\mathcal{O}(n^2 d)$ per layer. Summing over the $n_l$ layers gives the leading-order estimate
\begin{equation}
    C = \mathcal{O}\!\left(n_l\, n\, d^2\right) + \mathcal{O}\!\left(n_l\, n^2\, d\right),
    \label{eq:flops_per_step}
\end{equation}
where the first term dominates in the regime $d \gtrsim n$ and the second term dominates in the opposite regime $n \gtrsim d$. Subleading contributions from embedding, output projection, layer normalization, and nonlinearities are neglected, as they do not affect the leading scaling with $n_l$, $n$, and $d$.

We numerically verify the scaling of the forward cost $C$ with the input length $n$, as predicted by~\cref{eq:flops_per_step}. 
The exact number of FLOPs per forward pass is computed directly with the built-in FLOP-counting utilities provided by \texttt{JAX}~\cite{jax2018github}, applied to the compiled computational graph of the variational Ansatz.
~\cref{fig:complexity} shows that the cost contains a contribution that scales linearly with $n$ at smaller input lengths, together with a quadratic contribution that becomes dominant only at larger $n$. 
We remark that, for the system sizes considered in this work and currently accessible to numerical simulation methods~\cite{viteritti2026spatialattention}, the cost remains deeply in the regime of linear scaling, confirming the assumption underlying the analysis in~\cref{sec:size_consistency}.
This stands in sharp contrast to typical transformer applications in natural language processing and image classification, where sequence lengths are large enough that the quadratic cost of self-attention dominates~\cite{vaswani2017attention}.

\begin{figure}[t]
  \centering
  \includegraphics[width=1\columnwidth]{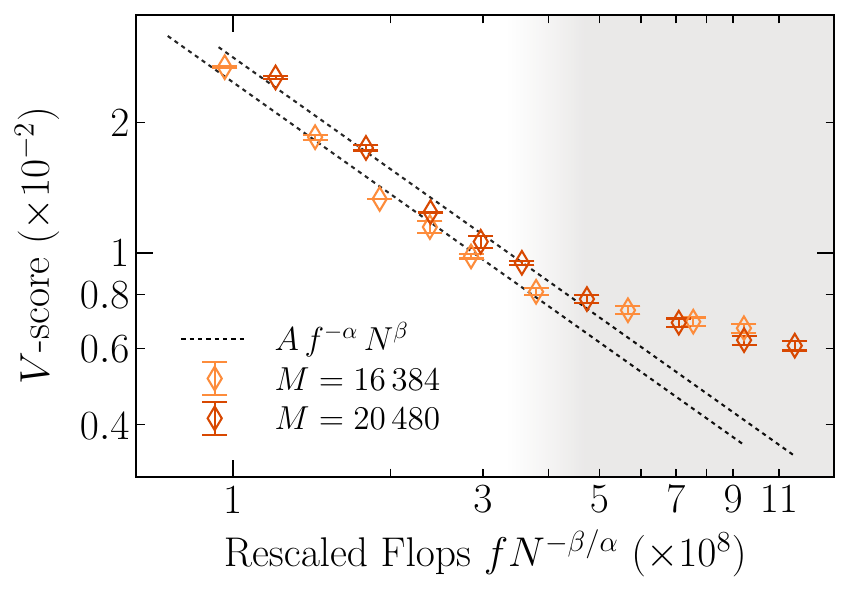}
\caption{\label{fig:samples}
$V$-score as a function of the rescaled compute $f N^{-\beta/\alpha}$ and corresponding power law fit for optimizations performed using two numbers of Monte Carlo samples, $M=16\,384$ and $M=20\,480$.
The system considered here is the square $J_1$-$J_2$ Heisenberg with $L=12$.
The regime where the scaling laws cease to apply is highlighted in grey.}
\end{figure}

The total training compute reported in the main text is then obtained by multiplying the per-configuration FLOP count $C$ by the number of Monte Carlo samples $M$ drawn at each optimization step and by the total number of optimization steps $N_{\mathrm{opt}}$,
\begin{equation}
    f = C \times M \times N_{\mathrm{opt}},
    \label{eq:total_flops}
\end{equation}
so that $f$ represents the cumulative number of forward-pass FLOPs expended during the entire ground-state optimization.

\subsection{Validity of Scaling Laws}\label{app:validity}
In this section, we discuss the regime of validity of the scaling laws. To probe this, we performed additional optimizations using architectures with more layers than those considered in the main analysis, up to $n_l=20$. As shown in \cref{fig:samples}, the $V$-score begins to saturate beyond $n_l=12$ layers. Such saturation may originate from intrinsic limitations of the architecture~\cite{viteritti2022,astrakhantsev2023}, from the finite number of Monte Carlo samples~\cite{sinibaldi2023unbiasing,gravina2025}, or from the optimization procedure itself. Any of these effects can compromise the applicability of the scaling laws in this large-compute regime.

We assess the role of sampling by comparing the results obtained with $M=16\,384$ samples, used in the main analysis, with those obtained using a larger sample size, namely $M=20\,480$. In \cref{fig:samples}, we show the $V$-score as a function of compute for increasing numbers of transformer layers and for both sample sizes. In the regime where the scaling law holds, the curves are described by a power law with the same exponent, indicating that our results are well converged with respect to the number of samples. By contrast, both sample sizes exhibit saturation in the large-compute regime, highlighted by the shaded region in \cref{fig:samples}, suggesting that this saturation is \textit{not} primarily controlled by sampling effects.

We therefore expect the main limitation of the observed behavior to arise from the stochastic optimization procedure. For very large architectures, the optimizer may fail to find sufficiently good minima, leading to a breakdown of the power-law regime. Improving the training strategy could therefore extend the validity of the scaling laws to larger architectures and higher training compute.

\begin{acknowledgments}
We thank Alessandro Laio, Shiwei Zhang, Anirvan Sengupta, Filippo Vicentini, Sebastian Goldt, and Rajah Nutakki for useful discussions. The Flatiron Institute is a division of the Simons Foundation. We acknowledge the CINECA award under the ISCRA initiative, for the availability of high-performance computing resources and support. LLV is supported by SEFRI under Grant No. MB22.00051 (NEQS - Neural Quantum). AS is supported by the Google PhD Fellowship 2025. The simulations presented in this work required a total computational budget of approximately $100\,000$ GPU hours on NVIDIA A100 GPUs.
\end{acknowledgments}

\bibliography{references}

\end{document}